\documentclass[12pt]{article}

\usepackage{subcaption}
\usepackage{tikz,pgfplots}
\usepackage{amsmath, amsfonts, amssymb}
\usepackage[margin=1in]{geometry}
\usepackage{authblk}
\usepackage{natbib}

\title{Velocity-Porosity Supermodel: A Deep Neural Networks based concept}

\author[1, 2]{D. Bhowmick}
\author[3]{D. K. Gupta}
\author[2]{S. Maiti}
\author[4]{U. Shankar}

\affil[1]{\small Centre for Mine Planning and Design Institute, Kanke Road, Ranchi 834008, Jharkhand, India}
\affil[2]{Dept. of Applied Geophysics, Indian Institute of Technology, ISM Dhanbad 826004, Jharkhand, India}
\affil[3]{Dept. of Precision and Microsystems Engineering, Delft University of Technology, Mekelweg 2, Delft 2628CD, The Netherlands}
\affil[4]{Dept. of Geophysics, Banaras Hindu University, Varanasi 221005, Uttar Pradesh, India}
\date{}

\begin{document}

\maketitle

\section{Introduction}

Rock physics models (RPMs) are used to estimate the elastic properties (\emph{e.g.} velocity, moduli) from the rock properties (\emph{e.g.} porosity, lithology, fluid saturation). However, the rock properties drastically vary for different geological conditions, and it is not easy to find a model that is applicable under all scenarios. There exist several empirical velocity-porosity transforms as well as first-principle-based models, however, each of these has its own limitations. For details on various RPMs, see the book by \cite{Mavko2009}. It is not so straight-forward to choose the correct RPM, and templates exist, which are overlapped with the log data to decide on the correct model.

In this work, we use deep machine learning and explore the concept of designing a supermodel that can be used for several different lithological conditions without any parameter tuning. In this paper, this test is restricted to only empirical velocity-porosity transforms, however, the future goal is to design a \emph{rock physics supermodel} that can be used on a variety of rock properties. In this paper, the idea is to combine the advantages of several existing empirical velocity-porosity transforms under a single framework, and design a velocity-porosity supermodel (VPS) using artificial neural networks (ANN) based deep learning.

\section{Method}

ANNs are algorithms which try to learn a pattern from provided datasets, similar to the way the human brain works. ANN has been used in various technical disciplines, and the field of geophysics is no exception (\emph{e.g.} \cite{Poulton1992}, \cite{McCormack1993}, \cite{Bhowmick2016}, among others). For our study, we look at two example cases. Case I attempts to combine the three empirical equations proposed by \cite{Raymer1980} into a single model. Case II considers empirical relations for 7 different lithologies \citep{Mavko2009} and combines them into one model.

\emph{Case I}: \cite{Raymer1980} proposed 3 relations between velocity and porosity, which are as follows.
\begin{align}
& 0 \leq \phi < 0.37, \quad \quad V_{p1} = (1 - \phi)^2V_{\text{ma}} + \phi V_\text{f}, \\
& \phi > 0.47, \qquad \qquad \frac{1}{\rho V_{p2}^2} = \frac{\phi}{\rho_f V_f^2} + \frac{1 - \phi}{\rho_{ma}V_{ma}^2}, \\ 
& 0.37 \leq \phi \leq 0.47, \quad \frac{1}{V_p} = \frac{\phi - 0.37}{0.1V_{p2}} + \frac{0.47 - \phi}{0.1V_{p1}},
\end{align}
where, $V_{ma}$ and $V_{f}$ denote matrix and fluid velocities, respectively. To be able to use the correct equation, one needs to study the porosity distribution for the provided dataset. Although here it is straightforward, we consider it as an excellent basic test for demonstrating the concept of a VPS. If the VPS can be trained to choose the correction equation for any provided dataset, it can certainly be extended to choose the correct empirical transform when several rock-properties and petrophysical properties need to be dealt with. To test the ANN model, datasets from NGHP-01-05 well site of KG Basin \citep{Shankar2013} and Mt. Elbert 01 well site of Alaska North Slope \citep{Rose2011} are used. The model is trained using 0.1 million training samples, where porosity, hydrate saturation and density are the inputs, and velocity is the target. The target $V_p$ values are calculated using the RHG equations, and the ANN model is trained to be able to match them. 

\emph{Case II}: Next, an ANN is trained able to choose the correct lithology, and accordingly determine $V_p$ and $V_s$ values based on one of the empirical transforms on which it has been trained. Table \ref{table_rpm1} lists lithologies and corresponding empirical transforms relating density ($\rho$), porosity ($\phi$), $V_p$ and $V_s$. The range of parameter values for which these transforms hold valid are reported in \mbox{Table \ref{table_rpm2}}. Using the information provided in the two tables, 0.175 milion training samples are generated and the ANN model is trained with $\phi$ and $\rho$ as inputs, and $V_p$ and $V_s$ as outputs. Synthetic data comprising the 7 lithologies is generated for test purpose. Once the ANN model has been trained, the goal is to accurately predict $V_p$ and $V_s$ values from $\phi$ and $\rho$ for the entire test dataset.
\begin{table}
\small
\caption{Empirical relations between $\phi$, $\rho$, $V_p$ and $V_s$ for different lithologies \citep{Mavko2009}}.
\vspace{-0.5em}
\begin{center}
\begin{tabular}{ l | c | c | c}
Lithology & $\phi - V_p$ & $\phi - V_s$ & $V_p - \rho$\\ 
 \hline
 Chalks & $V_p = 5.059\phi^2 - 8.505\phi + 5.128$ & $V_s = 2.766-2.933\phi$ & $\rho = 1.045 + 0.373V_p$ \\
 Dolomite & $V_p = 6.606 - 9.380\phi$ &  $V_S=3.581-4.719\phi$ & $\rho=1.843-0.137V_p$\\
 Sandstones & $V_p=4.944-5.201\phi$ & $V_s=2.981-3.484\phi$ & $\rho=1.569+0.195V_p$\\ 
Tight-gas sandstones & $V_p=4.868-3.836\phi$ & $V_s=3.149-1.703\phi$ & $\rho=1.96+0.117V_p$ \\
Limestone & $V_p=5.624-6.65\phi$ & $V_s=3.053-3.866\phi$ & $\rho=1.513+0.202V_p$ \\
High-porosity sandstones & $V_p=4.303-2.227\phi$ & $V_s=2.486-1.626\phi$ & $\rho=1.450+0.219V_p$ \\ 
Poorly consolidated sandstones & $V_p=3.774-3.414\phi$ & $V_s=2.100-2.424\phi$ & $\rho=1.498+0.224V_p$ \\
\end{tabular}
\end{center}
\label{table_rpm1}
\end{table}

\begin{table}
\caption{Range of values for the applicability of empirical relations presented in Table \ref{table_rpm1}.}
\vspace{0.5em}
\begin{center}
\begin{tabular}{ l | c | c | c | c}
Lithology & $\phi$ & $\rho$ (gcm$^{-3}$) & $V_p$ (ms$^{-1}$) & $V_s$ (ms$^{-1}$)\\ 
 \hline
 Chalks & 0.10 - 0.75 & 1.43 - 2.57 & 1.53 - 4.30 & 1.59 - 2.51 \\
 Dolomite & 0.00 - 0.32 & 2.27 - 2.84 & 3.41 - 7.02 & 2.01 - 3.64 \\
 Sandstones & 0.04 - 0.30 & 2.09 - 2.64 & 3.13 - 5.52 & 1.73 - 3.60\\ 
Tight-gas sandstones & 0.01 - 0.14 & 2.26 - 2.67 & 3.81 - 5.57 & 2.59 - 3.50 \\
Limestone & 0.03 - 0.41 & 2.00 - 2.65 & 3.39 - 5.79 & 1.67 - 3.04 \\
High-porosity sandstones & 0.02 - 0.32 & 2.12 - 2.69 & 3.46 - 4.79 & 1.95 - 2.66 \\ 
Poorly consolidated sandstones & 0.22 - 0.36 & 2.01 - 2.23 & 2.43 - 3.14 & 1.21 - 1.66 \\
\end{tabular}
\end{center}
\label{table_rpm2}
\end{table}

\section{Results}

\begin{figure} 
\begin{center}
    \begin{tikzpicture}[scale = 0.4]
        \begin{axis}[%
        		thick,
            axis x line=bottom,
            axis y line=left,
            legend pos = north west,
            legend style = {font=\LARGE},
            ymin = 1440,
            ymax = 1580,
            xlabel = {Log data samples},
            xlabel style = {xshift = -1950pt, yshift= -5 pt},
            ylabel = {$V_p$ (in ms$^{-1}$)},
            xscale = 5,
            label style = {font=\LARGE},
            tick label style = {font=\LARGE},
            ylabel style={yshift=30pt},
            y tick label style={/pgf/number format/.cd,%
            set thousands separator={}},
            ]

            \addplot[mark=none, solid, blue, very thick] table[x= sno, y = Vtarget] {KG_RHG_NET_COMB_9_15_9.TXT};
            \addplot[mark=none, solid, red, very thick] table[x= sno, y = Vobserved] {KG_RHG_NET_COMB_9_15_9.TXT};
			\addlegendentry{Target data}
            \addlegendentry{Supermodel output}
        \end{axis}
    \end{tikzpicture}
    \begin{tikzpicture}[scale = 0.4]
        \begin{axis}[%
        		thick,
            axis x line=bottom,
            axis y line=left,
            legend pos = north west,
            ymin = -0.006,
            ymax = 0.004,
            xlabel = {Log data samples},
            xlabel style = {xshift = -1950pt, yshift= -5 pt},
            ylabel = {Normalized error},
            xscale = 5,
            label style = {font=\LARGE},
            tick label style = {font=\LARGE},
            ylabel style={yshift=10pt}
                        ]

            \addplot[mark=none, solid, blue, very thick] table[x=sno, y = normerror] {KG_RHG_NET_COMB_9_15_9.TXT};
        \end{axis}
    \end{tikzpicture}
\end{center}
\vspace{-1em}
\caption{Target $V_p$ values (calculated using RHG equation) and $V_p$ values obtained using a trained velocity-porosity supermodel (top) for the NGHP-01-05 site of Krishna-Godavari basin, India, and the normalized error values (bottom). The error values have been normalized using the largest $V_p$ value in the entire target set.}
\label{fig_rpmresult1}	
\end{figure}
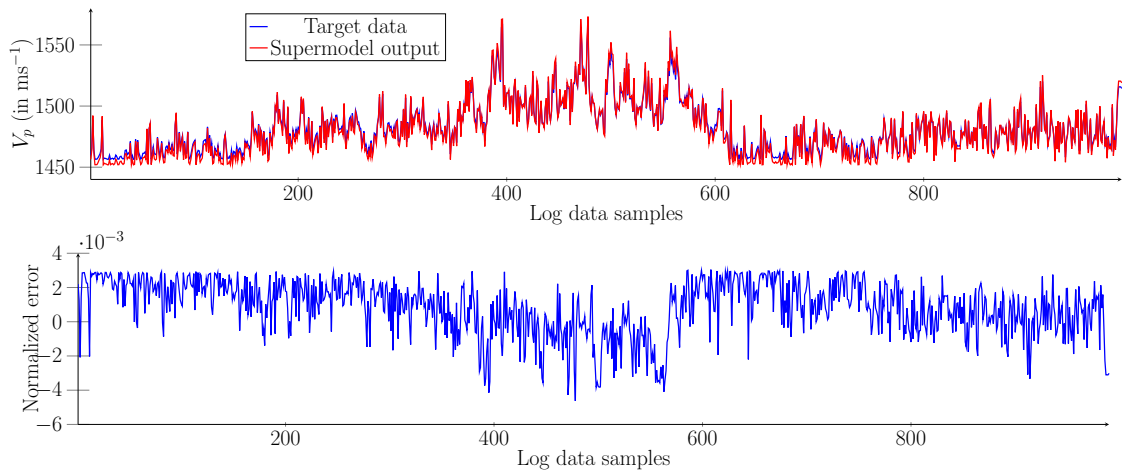

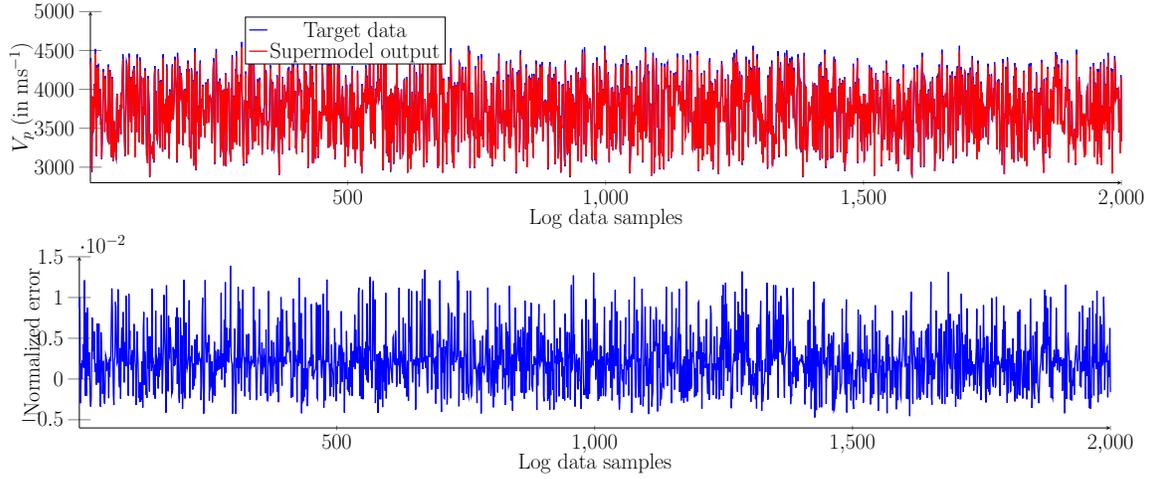
\begin{figure} 
\begin{center}
    \begin{tikzpicture}[scale = 0.4]
        \begin{axis}[%
        		thick,
            axis x line=bottom,
            axis y line=left,
            legend pos = north west,
            legend style = {font=\LARGE},
            ymin = 2800,
            ymax = 5000,
            xlabel = {Log data samples},
            xlabel style = {xshift = -1950pt, yshift= -5 pt},
            ylabel = {$V_p$ (in ms$^{-1}$)},
            xscale = 5,
            label style = {font=\LARGE},
            tick label style = {font=\LARGE},
            ylabel style={yshift=30pt},
            y tick label style={/pgf/number format/.cd,%
            set thousands separator={}},
            ]

            \addplot[mark=none, solid, blue, very thick] table[x= sno, y = Vtarget] {ME_RHG_NET_COMB_9_15_9.TXT};
            \addplot[mark=none, solid, red, very thick] table[x= sno, y = Vobserved] {ME_RHG_NET_COMB_9_15_9.TXT};
			\addlegendentry{Target data}
            \addlegendentry{Supermodel output}
        \end{axis}
    \end{tikzpicture}
    \begin{tikzpicture}[scale = 0.4]
        \begin{axis}[%
        		thick,
            axis x line=bottom,
            axis y line=left,
            legend pos = north west,
            ymin = -0.006,
            ymax = 0.015,
            xlabel = {Log data samples},
            xlabel style = {xshift = -1950pt, yshift= -5 pt},
            ylabel = {Normalized error},
            xscale = 5,
            label style = {font=\LARGE},
            tick label style = {font=\LARGE},
            ylabel style={yshift=10pt}
                        ]

            \addplot[mark=none, solid, blue, very thick] table[x=sno, y = normerror] {ME_RHG_NET_COMB_9_15_9.TXT};
        \end{axis}
    \end{tikzpicture}
\end{center}
\vspace{-1em}
\caption{Target $V_p$ values (calculated using RHG equation) and $V_p$ values obtained using a trained velocity-porosity supermodel (top) for the Mt. Elbert 01 site of Alaska North Slope, and the normalized error values (bottom). The error values have been normalized using the largest $V_p$ value in the entire target set.}
\label{fig_rpmresult2}	
\end{figure}

\vspace{-0.5em}
The velocity-porosity supermodels (VPSs) for the two cases have been trained using several different neural nets. Table \ref{table_rpm3} shows the number of hidden layers and the respective normalized root mean-squared errors (denoted by $\epsilon(\cdot)$ functional) obtained for the two cases. For Case I, relatively shallow networks are found to be sufficient, and with 3 hidden layers comprising 9, 15 and 9 neurons, the error is already very small. Fig. \ref{fig_rpmresult1} shows the $V_p$ log obtained using Raymer-Hunt-Gardner transform and the one obtained from the trained VPS using 3 hidden layers of 9, 15 and 9 neurons. From the error plot, it is observed that our VPS can very well predict the $V_p$ values for this data. Similarly, Fig. \ref{fig_rpmresult2} shows the target and VPS predicted $V_p$ values for Mt. Elbert site. For this data as well, the trained VPS can very accurately predict $V_p$ values as seen from its error plot.

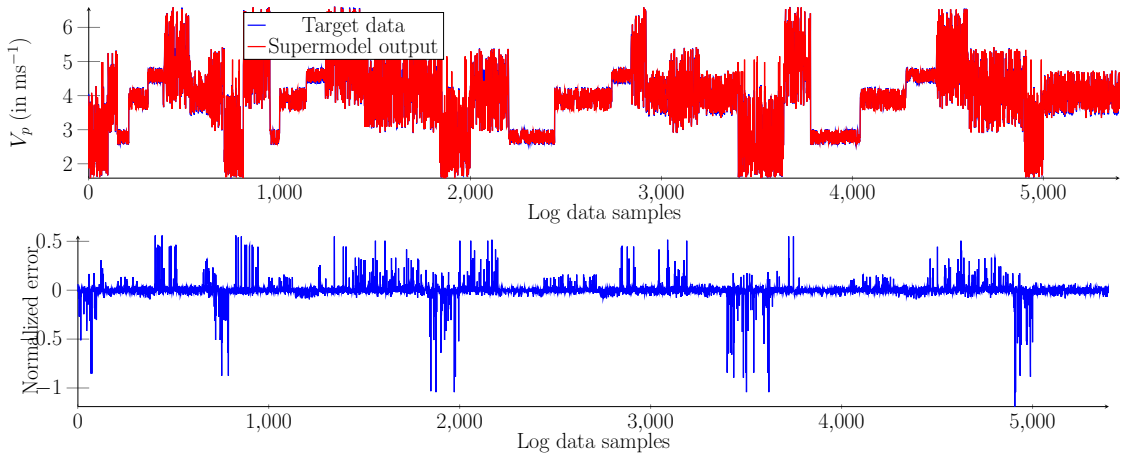
\begin{figure} 
\begin{center}
    \begin{tikzpicture}[scale = 0.4]
        \begin{axis}[%
        		thick,
            axis x line=bottom,
            axis y line=left,
            legend pos = north west,
            legend style = {font=\LARGE},
            xlabel = {Log data samples},
            xlabel style = {xshift = -1950pt, yshift= -5 pt},
            ylabel = {$V_p$ (in ms$^{-1}$)},
            xscale = 5,
            label style = {font=\LARGE},
            tick label style = {font=\LARGE},
            ylabel style={yshift=30pt},
            y tick label style={/pgf/number format/.cd,%
            set thousands separator={}},
            ]

            \addplot[mark=none, solid, blue, very thick] table[x= sno, y = Vpt] {net_7_15_21_15_7.TXT};
            \addplot[mark=none, solid, red, very thick] table[x= sno, y = Vpo] {net_7_15_21_15_7.TXT};
			\addlegendentry{Target data}
            \addlegendentry{Supermodel output}
        \end{axis}
    \end{tikzpicture}
    \begin{tikzpicture}[scale = 0.4]
        \begin{axis}[%
        		thick,
            axis x line=bottom,
            axis y line=left,
            legend pos = north west,
            xlabel = {Log data samples},
            xlabel style = {xshift = -1950pt, yshift= -5 pt},
            ylabel = {Normalized error},
            xscale = 5,
            label style = {font=\LARGE},
            tick label style = {font=\LARGE},
            ylabel style={yshift=10pt}
                        ]

            \addplot[mark=none, solid, blue, very thick] table[x=sno, y = Vperror] {net_7_15_21_15_7.TXT};
        \end{axis}
    \end{tikzpicture}
\end{center}
\vspace{-1em}
\caption{Target $V_p$ values and $V_p$ values obtained using a trained velocity-porosity supermodel (top) for the synthetic well data and the normalized error values (bottom). The error values have been normalized using the largest $V_p$ value in the entire target set.}
\label{fig_rpmresult3}	
\end{figure}

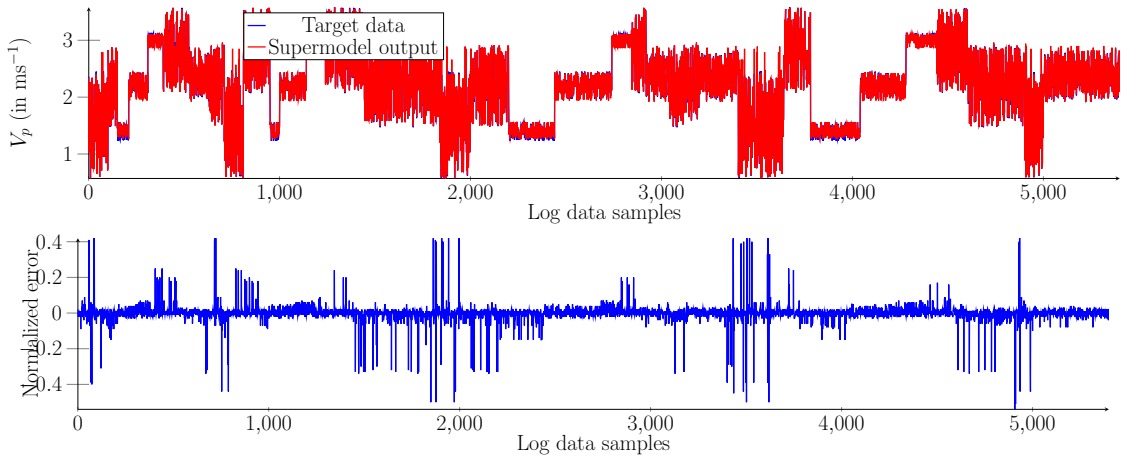
\begin{figure} 
\begin{center}
    \begin{tikzpicture}[scale = 0.4]
        \begin{axis}[%
        		thick,
            axis x line=bottom,
            axis y line=left,
            legend pos = north west,
            legend style = {font=\LARGE},
            xlabel = {Log data samples},
            xlabel style = {xshift = -1950pt, yshift= -5 pt},
            ylabel = {$V_p$ (in ms$^{-1}$)},
            xscale = 5,
            label style = {font=\LARGE},
            tick label style = {font=\LARGE},
            ylabel style={yshift=30pt},
            y tick label style={/pgf/number format/.cd,%
            set thousands separator={}},
            ]

            \addplot[mark=none, solid, blue, very thick] table[x= sno, y = Vst] {net_7_15_21_15_7.TXT};
            \addplot[mark=none, solid, red, very thick] table[x= sno, y = Vso] {net_7_15_21_15_7.TXT};
			\addlegendentry{Target data}
            \addlegendentry{Supermodel output}
        \end{axis}
    \end{tikzpicture}
    \begin{tikzpicture}[scale = 0.4]
        \begin{axis}[%
        		thick,
            axis x line=bottom,
            axis y line=left,
            legend pos = north west,
            xlabel = {Log data samples},
            xlabel style = {xshift = -1950pt, yshift= -5 pt},
            ylabel = {Normalized error},
            xscale = 5,
            label style = {font=\LARGE},
            tick label style = {font=\LARGE},
            ylabel style={yshift=10pt}
                        ]

            \addplot[mark=none, solid, blue, very thick] table[x=sno, y = Vserror] {net_7_15_21_15_7.TXT};
        \end{axis}
    \end{tikzpicture}
\end{center}
\vspace{-1em}
\caption{Target $V_s$ values and $V_s$ values obtained using a trained velocity-porosity supermodel (top) for the synthetic well data and the normalized error values (bottom). The error values have been normalized using the largest $V_s$ value in the entire target set.}
\label{fig_rpmresult4}	
\end{figure}

From the error values reported in Table \ref{table_rpm3}, it is seen that for Case II, even with 5 hidden layers, $V_p$ and $V_s$ cannot be predicted to a higher degree of accuracy. Fig. \ref{fig_rpmresult3} shows the target and predicted $V_p$ values, and the respective error plot obtained using 5 hidden layers with 7, 15, 21, 15 and 7 neurons. From the error plot, we observe that there are certain zones where the error is relatively very high. From cross-validation, it was observed that all these zones correspond to chalks. The reason for the trained VPS to perform poorly in these zones is that the relation between $\phi$ and $V_p$ in the region is quadratic, compared to linear relation in all other chosen lithologies. For ANN, fitting a quadratic is more difficult compared to linear relations, due to which it fits poorly for chalk. However, this is certainly not a problem, and with a larger network, this issue can be eliminated. From Fig. \ref{fig_rpmresult4}, we see that the fit for $V_s$ is similar to that of $V_p$.

\begin{table}
\caption{Neural nets and the respective normalized root-mean-squared error values for the two test cases.}
\small
\begin{center}
\begin{tabular}{ c | c  | c || c | c | c}
\multicolumn{3}{c | }{Case I} & \multicolumn{3}{ | c}{Case II}\\
\hline
Network size & $\epsilon(V_p)$, NGHP & $\epsilon(V_p)$, Mt. Elbert & Network size & $\epsilon(V_p)$ & $\epsilon(V_s)$\\ 
 \hline
5 & 0.04460 & 0.10887 & 4 & 0.19066 & 0.12267\\
5, 7, 5 & 0.00898 & 0.00999 & 3, 5, 3 & 0.17081 & 0.10258\\
5, 9, 5 & 0.00946 & 0.00853 & 4, 7, 4 & 0.13804 & 0.08055\\
7, 11, 7 & 0.00518 & 0.00633 & 4, 7, 10, 7, 4 & 0.11096 & 0.06604\\
9, 15, 9 & 0.00178 & 0.00426 & 7, 15, 21, 15, 7 & 0.10012 & 0.05700\\
\end{tabular}
\end{center}
\label{table_rpm3}
\end{table}

\section{Conclusions}

This paper explores the concept of designing a velocity-porosity supermodel using deep neural networks. Two test examples are demonstrated in this paper, and it is shown that ANN based supermodel can learn the behavior of the existing velocity-porosity models. Irrespective of the lithology, the trained supermodel can be used on any novel datasets as long as it has been trained sufficiently. The futuristic idea that motivates this study is to design a rock-physics supermodel which can be considered as a universal approximator for various rock properties.
 
 \bibliography{cph18_start}

\end{document}